\documentclass[pdflatex,sn-mathphys-num]{sn-jnl}

\usepackage{graphicx}%
\usepackage{multirow}%
\usepackage{amsmath,amssymb,amsfonts}%
\usepackage{amsthm}%
\usepackage{mathrsfs}%
\usepackage[title]{appendix}%
\usepackage{xcolor}%
\usepackage{textcomp}%
\usepackage{manyfoot}%
\usepackage{booktabs}%
\usepackage{algorithm}%
\usepackage{algorithmicx}%
\usepackage{algpseudocode}%
\usepackage{listings}%

\theoremstyle{thmstyleone}%
\theoremstyle{thmstyletwo}%

\theoremstyle{thmstylethree}%

\begin{document}

\title[Golden Tonnetz]{Golden Tonnetz}


\author*[1]{\fnm{Yusuke} \sur{Imai}}\email{imai@isi.imi.i.u-tokyo.ac.jp}

\affil*[1]{\orgdiv{Graduate School of Information Science and Technology}, \orgname{The University of Tokyo}, \orgaddress{\street{7-3-1 Hongo}, \city{Bunkyo}, \postcode{113-8656}, \state{Tokyo}, \country{Japan}}}

\abstract{Musical concepts have been represented by geometry with tones. 
For example, in the chromatic circle, the twelve tones are represented by twelve points on a circle, and in Tonnetz, the relationships among harmonies are represented by a triangular lattice.
Recently, we have shown that several arrangements of tones on the regular icosahedron can be associated with chromatic scales, whole-tone scales, major tones, and minor tones through the golden ratio.
Here, we investigate another type of connection between music and the golden ratio.
We show that there exists an arrangement of 7 tones on a golden triangle that can represent a given major/minor scale and its tonic, dominant, and subdominant chords by golden triangles.
By applying this finding, we propose ``golden Tonnetz" which represents all the major/minor scales and triads by the golden triangles or gnomons and also represents relative, parallel, and leading-tone exchange transformations in Neo-Riemannian theory by transformations among the golden triangles and gnomons.}




\maketitle

\section{Introduction}\label{sec1}

Musical concepts such as harmony and scale have been characterized by geometry.
For example, the chromatic circle contained twelve tones in a circle, and harmonies and scales are represented by figures obtained by connecting tones in the circle \cite{mccartin1998prelude}.
In addition, Tonnetz characterizes the harmonic relationships by a triangular lattice \cite{Euler1739}, and various generalizations of Tonnetz have been proposed \cite{catanzaro2011generalized,mohanty20225,rietsch2024generalizations,boland2025three}.
Recently, we found that the icosahedron connects various musical concepts, including the chromatic scale, whole tone scales, major and minor triads, and Gregorian modes through the golden ratio \cite{imai2021general1,imai2021general2,imai2021general3}.
In this paper, we find another example of characterizing musical concepts through the golden ratio by utilizing the regular pentagon.
This paper studies extensions of Tonnetz using golden triangles and gnomons, where the golden triangles (gnomons) are isosceles triangles whose length ratio of the longest side to the shortest side is given by the golden ratio without obtuse angles (with an obtuse angle), as shown in Fig.~\ref{fig1}a).
The simple extension is the replacement of the equilateral triangle in Tonnetz by the golden triangles/gnomons.
Here, we propose another kind of extension of Tonnetz by using golden triangles and gnomons that involves not only the major/minor triads but also major/minor scales and Gregorian modes.

\section{Golden triangle and major scale}\label{sec2}
\begin{figure}[t]
\centering
\includegraphics[width=\textwidth]{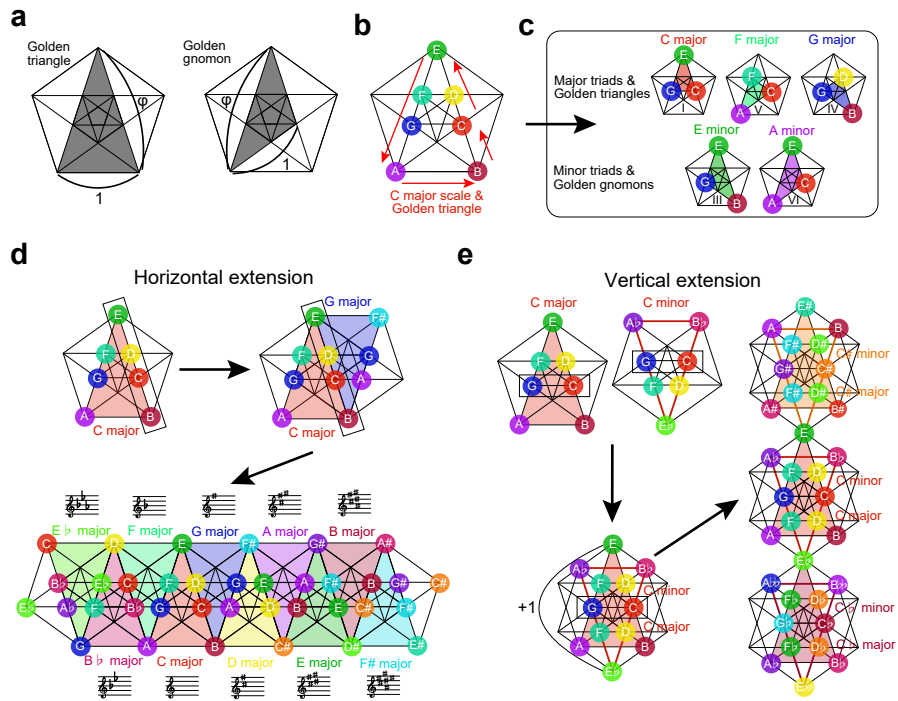}
\caption{\textbf{a}, Golden triangle and gnomon . \textbf{b}, The base figure where the $C$ major scale on the golden triangle. \textbf{c}, Triads represented by golden triangles or gnomons in the base figure. \textbf{d}, Horizontal extension of the base figure. \textbf{e}, Vertical extension of the base figure.}\label{fig1}
\end{figure}

We first found that there exists an arrangement of 7 tones on a given golden triangle (Fig.~\ref{fig1}b) that satisfies the following two conditions.

(1) The neighboring tones in $C$ major scale neighbor in terms of the graph topology (e.g., $D$ is linked to $C$ and $E$) as shown in Fig.~\ref{fig1}b.

(2) The chords I (tonic), III, IV (subdominant), V (dominant), VI are represented by the golden triangle or gnomon (Fig.~\ref{fig1}c).

We call this figure the base figure, and there is a way to extend the base figure horizontally and vertically.

In Fig.~\ref{fig1}d, we extend the base figure horizontally.
Because the right part of the base figure contains $B$, $C$, $D$, $E$, then, one can connect the $G$ major scale version of the base figure to the base figure because the $G$ major scale version also contains $B$, $C$, $D$, $E$ in the right part (see the tones in a rectangular in Fig.~\ref{fig1}d).
Similarly, one can extend the base figure horizontally so that the major scale on the golden triangle one right has one more sharp than the major scale in the original golden triangle. ($\to E\flat\to B\flat \to F \to C \to G \to D \to A \to E \to B \to F\sharp \to$).

The vertical extension of the base figure is shown in Fig.~\ref{fig1}\textbf{e}.
We first prepare the base figure and the $C$ minor version of the base figure.
Because the figures share $C$ and $G$, the figures can be connected so that $C$ and $G$ overlap.
The resulting figures (bottom left figure of Fig.~\ref{fig1}e) have $E$ at the top and $E\flat$ at the bottom.
Because $E$ is 1 semitone higher than $E\flat$, one can extend the base figure vertically so that the major/minor scale in the golden triangles above one has one more sharp than the major/minor scale in the original golden triangle ($\to C\flat$ major/minor $\to C$ major/minor $\to C\sharp$ major/minor) as shown in the right figure of Fig.~\ref{fig1}\textbf{e}.

Other choices of the base figures and their extensions are discussed in Figs.~\ref{fig5}-\ref{fig11}.

\section{Golden Tonnetz}\label{sec3}
\begin{figure}[t]
\centering
\includegraphics[width=\textwidth]{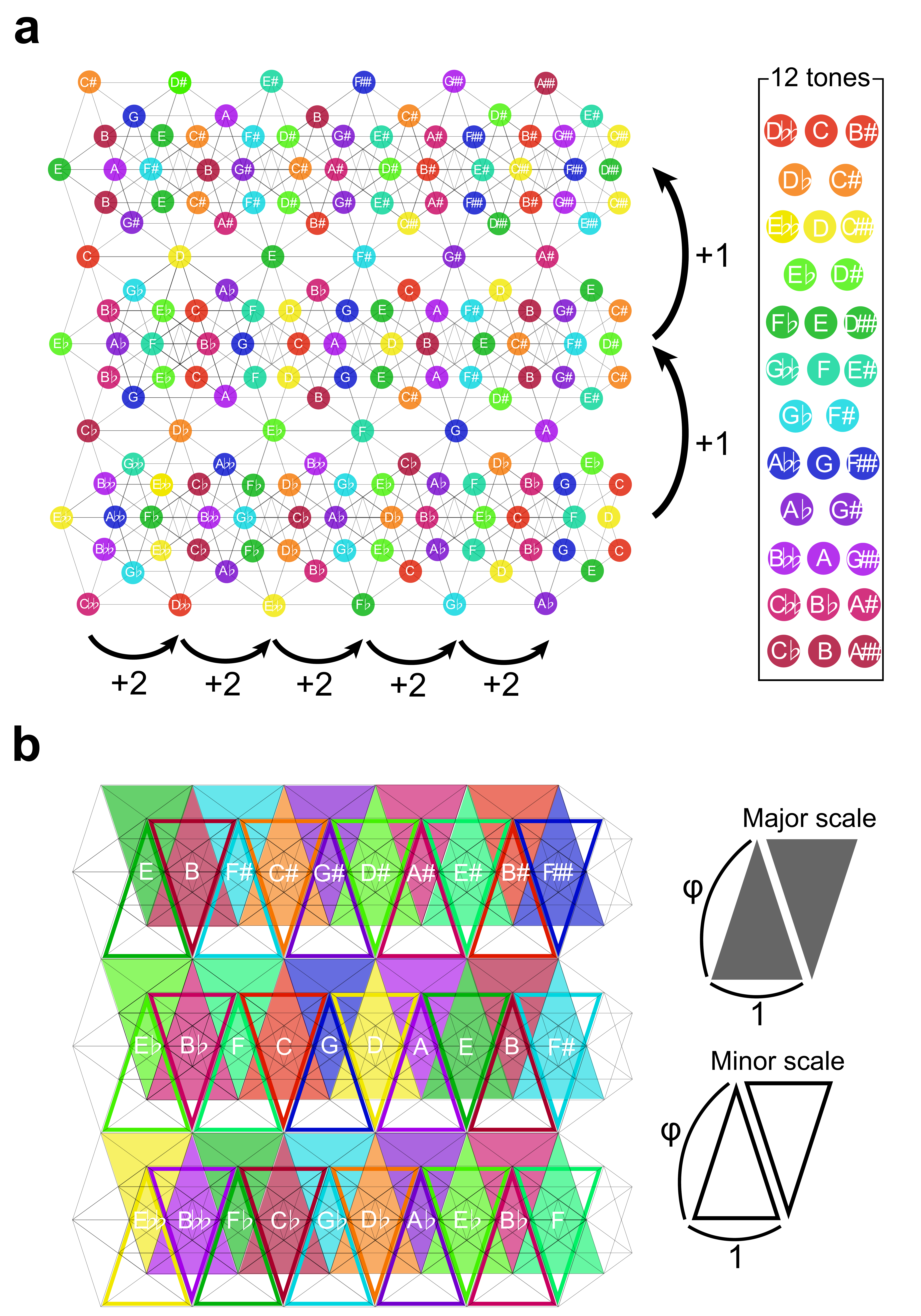}
\caption{\textbf{a}, Golden Tonnetz obtained by horizontally and vertically extending the base figure (Fig.~\ref{fig1}\textbf{b}). \textbf{b}, Major and minor scales represented by golden triangles on the golden Tonnetz. \textbf{c}, Basis of the golden Tonnetz.}\label{fig2}
\end{figure}

By applying the horizontal and vertical extension to the base figure infinitely, we obtain the lattice of the golden triangles with tones (Fig.~\ref{fig2}a). 
We call this figure the golden Tonnetz because this figure has a structure similar to Tonnetz.
By definition, all major/minor scales and triads are represented by golden triangles in the golden Tonnetz (Fig.~\ref{fig2}b).
Figure \ref{fig2}c shows the basis of the golden Tonnetz.
The golden Tonnetz is constructed by translating the basic figure, whose vertices correspond to tones, such that a horizontal shift by one unit to the right (left) raises (lowers) each tone by a perfect fifth, and a vertical shift upward (downward) raises (lowers) it by a semitone.
The notes contained in the right-hand figure of the basic shapes are the notes contained in the left-hand figure raised by a fifth. Furthermore, the scale on one golden triangle in each figure is the relative of the scale on the other golden triangle.
Additionally, the tonic, dominant, and subdominant chords of each scale are also represented by golden triangles.

\begin{figure}[t]
\centering
\includegraphics[width=\textwidth]{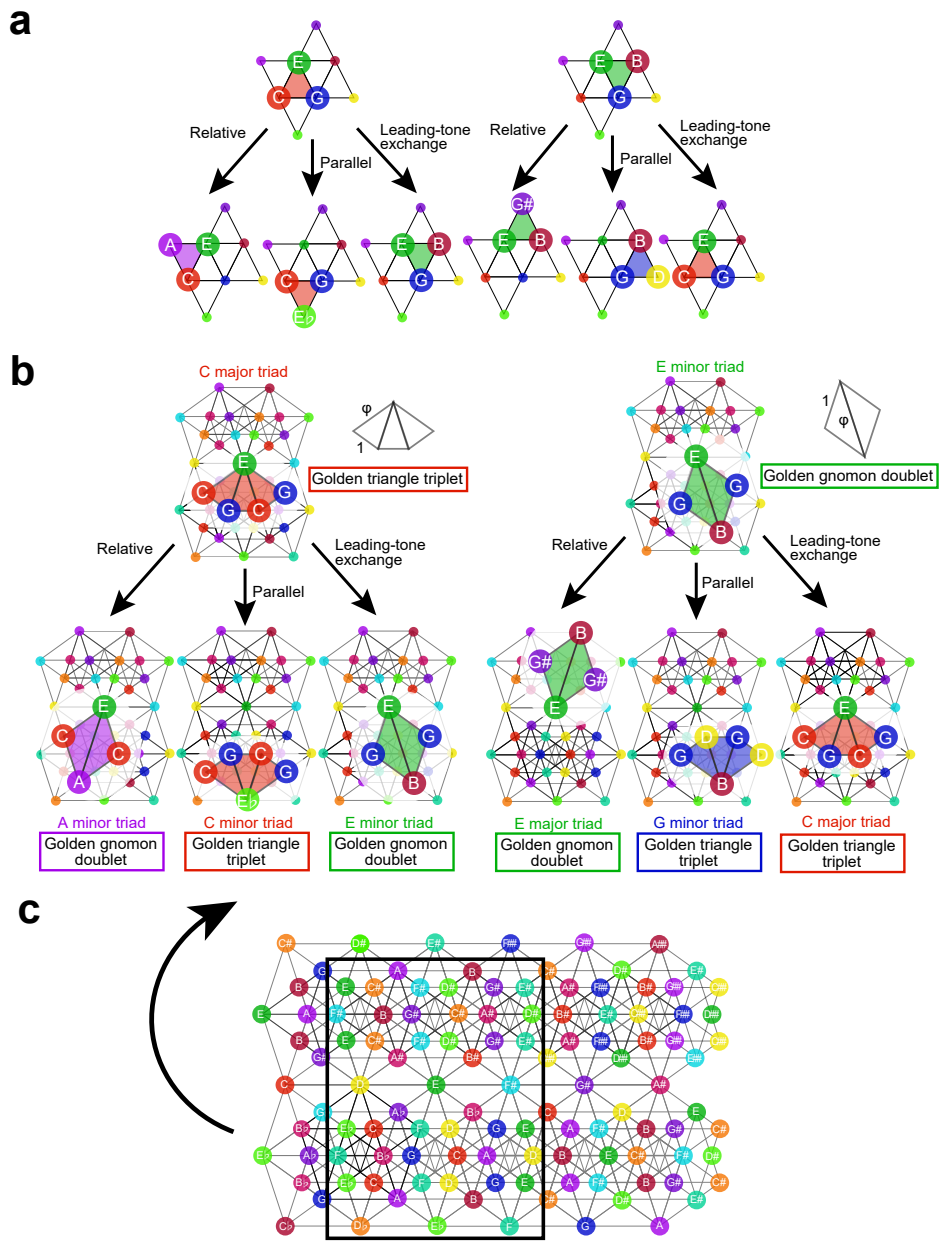}
\caption{\textbf{a}, Relative, parallel, and Leading-tone exchange transformations by Tonnetz, \textbf{b}, by golden Tonnetz. \textbf{c}, The part of golden Tonnetz used in \textbf{b}.}\label{fig3}
\end{figure}

In Tonnetz, all major/minor triads are represented by equilateral triangles.
The relative ($CEG\to CEA$, $EGB\to EG\sharp B$), parallel ($CEG\to CE\flat G$, $EGB\to DGB$), and leading-tone exchange ($CEG\to BEG$, $EGB\to EGC$) transformations are represented by the transformations between the neighboring equilateral triangles (Fig.~\ref{fig3}a).

In the golden Tonnetz, all major/minor triads are represented by golden triangle triplets or gnomon doublets.
Figure \ref{fig3}b summarizes the relative, parallel, and leading-tone exchange transformations in the golden Tonnetz.
First, $CEG$ are represented by the golden triangle triplet.
Then, one can represent the relative transformation for the $C$ major triads by the transformation of the golden triangle triplet with the $C$ major triads into the neighboring golden gnomon doublet with the $A$ minor triads.
In addition, one can represent the parallel transformation for the $C$ major triads by the transformation of the golden triangle triplet with the $C$ major triads into the neighboring golden triangle triplet with the $C$ minor triads.
Furthermore, one can represent the leading-tone exchange transformation for the $C$ major triads by the transformation of the golden triangle triplet with the $C$ major triads into the neighboring golden gnomon doublet with the $E$ minor triads.
Similarly, the relative, parallel, and leading-tone exchange transformation for the $E$ minor triad can be obtained by the transformations between the golden gnomon doublet representing the $E$ minor triads and the golden gnomon doublet representing the $E$ major triad and the golden triangle triplet representing the $G$ major triad and the $C$ major triad.

\begin{figure}[t]
\centering
\includegraphics[width=\textwidth]{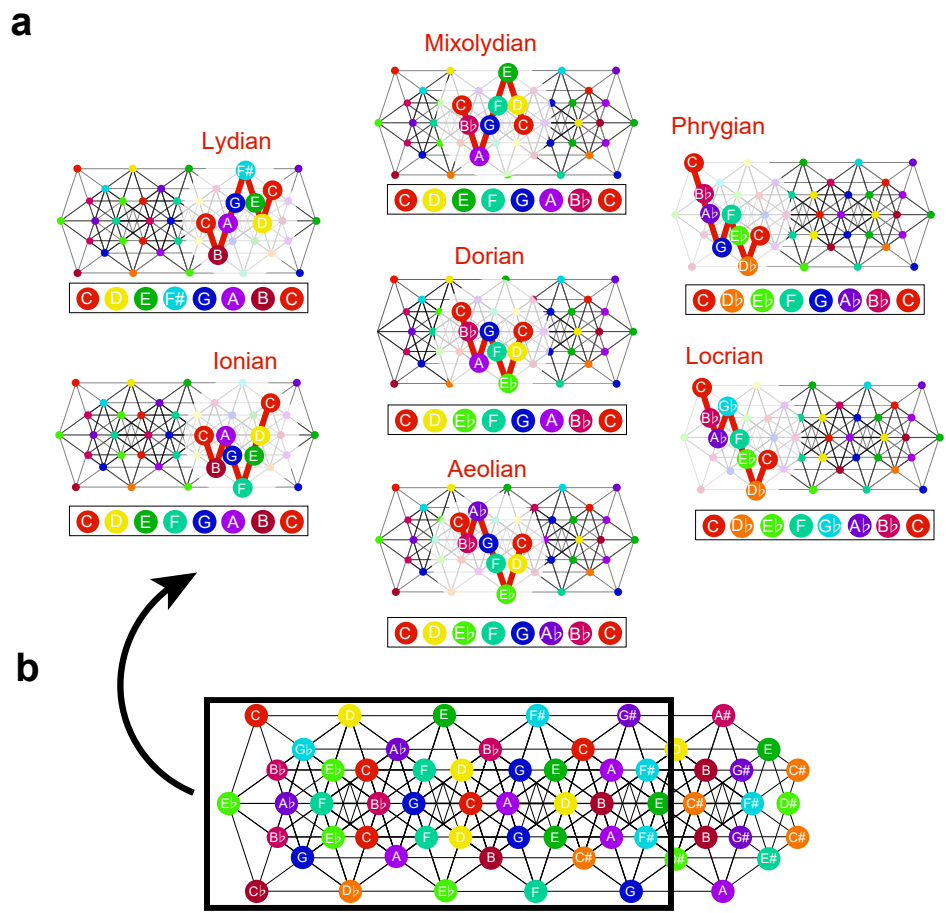}
\caption{\textbf{a}, Gregorian modes (Lydian, Ionian, Mixolydian, Dorian, Aeolian, Phrygian, and Locrian modes) on golden Tonnetz. \textbf{b}, The part of golden Tonnetz used in \textbf{a}.}\label{fig4}
\end{figure}

\begin{figure}[t]
\centering
\includegraphics[width=\textwidth]{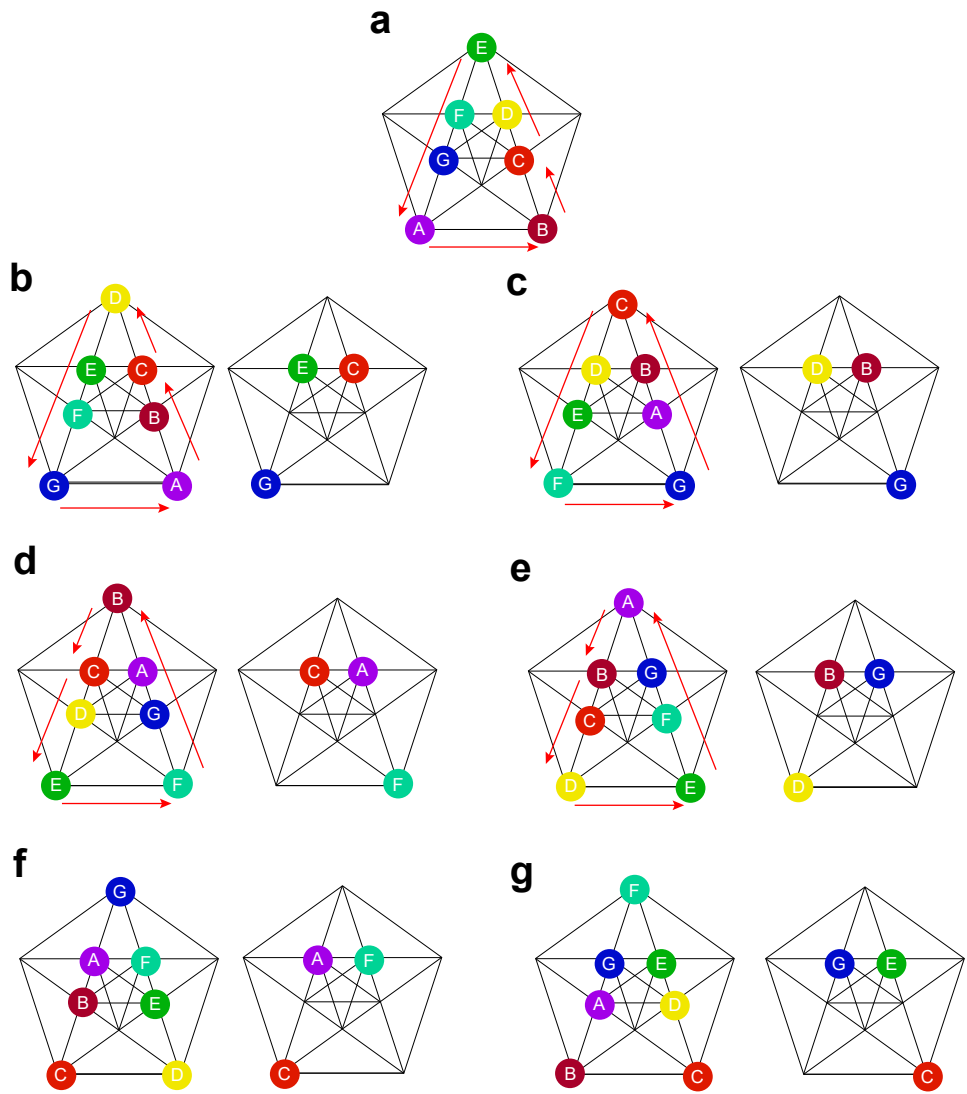}
\caption{\textbf{a}--\textbf{g}, Seven kinds of arrangement of $C$ major scale on the golden triangle where \textbf{a} is shown in Fig.~\ref{fig1}. Only \textbf{a} is an arrangement consistent with the golden ratio. For example, in the arrangement of \textbf{b}, the triad $CEG$ (tonic) is not represented by a golden triangle or gnomon.} \label{fig5}
\end{figure}

\begin{figure}[t]
\centering
\includegraphics[width=\textwidth]{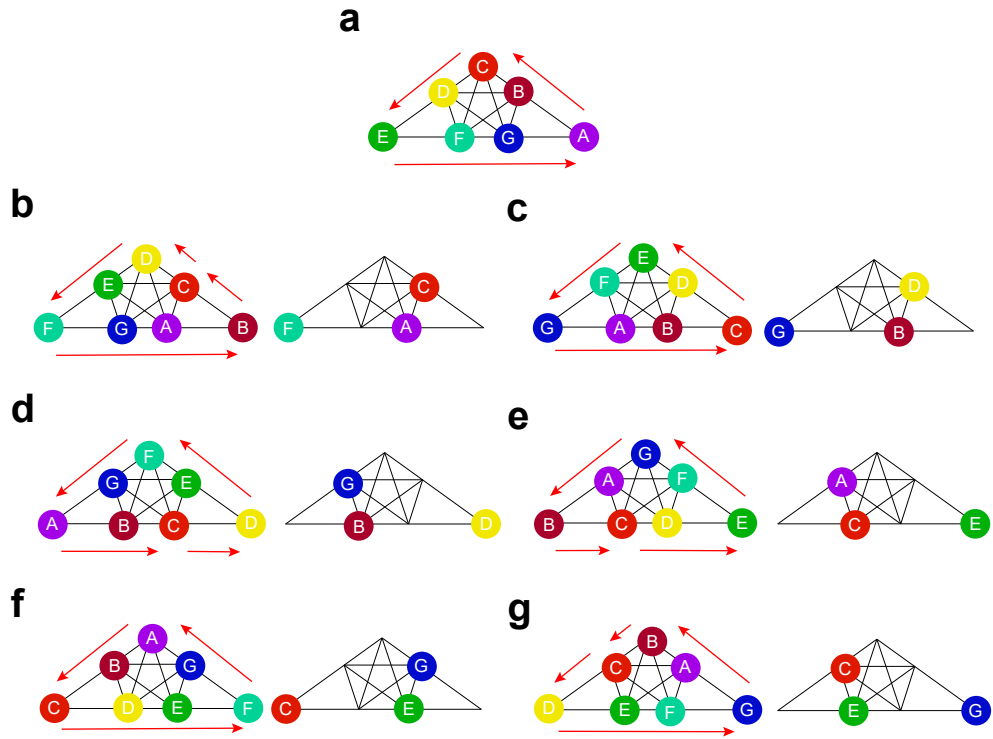}
\caption{\textbf{a}--\textbf{g}, Seven kinds of arrangement of $C$ major scale on the golden gnomon.
Only \textbf{a} is an arrangement consistent with the golden ratio, but not consistent with the horizontal or vertical extension.}\label{fig6}
\end{figure}

\begin{figure}[t]
\centering
\includegraphics[width=0.7\textwidth]{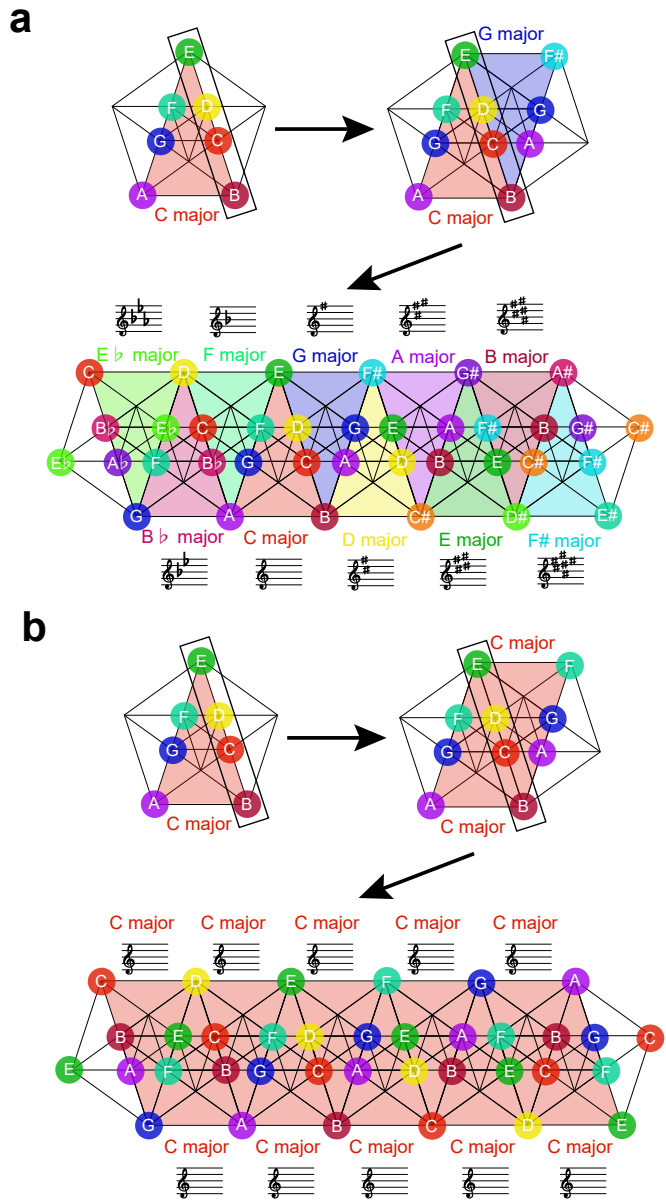}
\caption{\textbf{a}--\textbf{b}, Two kinds of the horizontal extension. \textbf{a} is shown in  Fig.~\ref{fig1}. The major scales other than the $C$ major scale cannot be represented by a golden triangle in \textbf{b}. All the arrangements shown in Fig.~\ref{fig5} are involved in \textbf{b}. There exist only two kinds of horizontal extension because major scales including $BCDE$ is only the $C$ and $G$ major scales.}\label{fig7}
\end{figure}

\begin{figure}[t]
\centering
\includegraphics[width=0.7\textwidth]{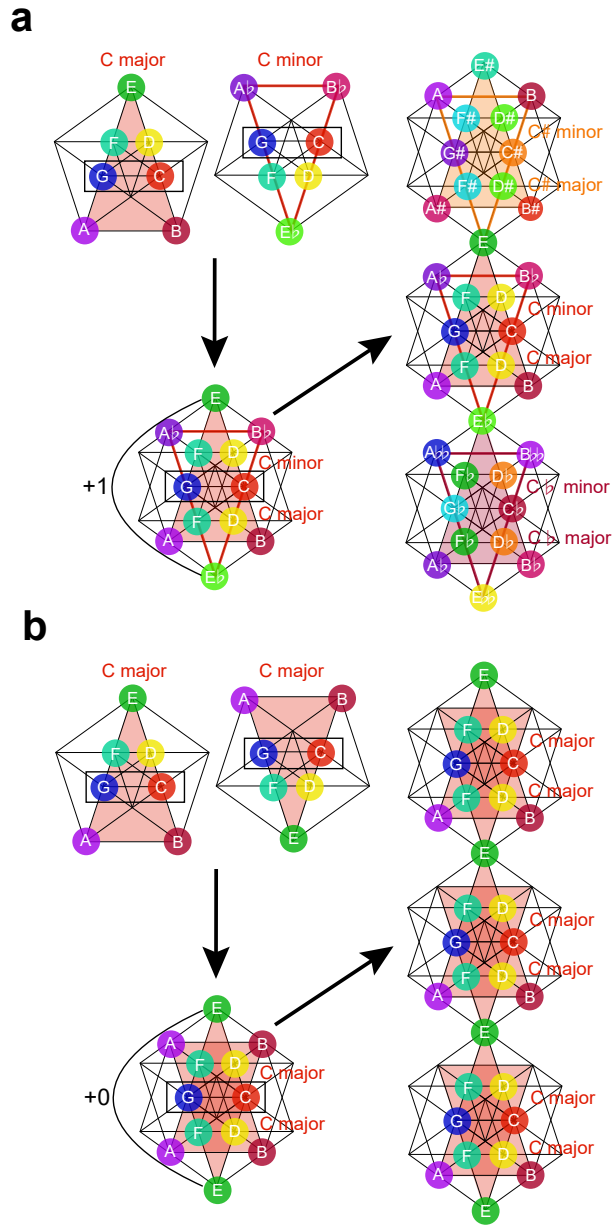}
\caption{\textbf{a}--\textbf{b}, Two kinds of the vertical extension. \textbf{a} is shown in  Fig.~\ref{fig1}. Any minor scales cannot be represented by a golden triangle in \textbf{b}. There exist six kinds of horizontal extension because major scales including $CG$ is the $C/A$ and $G/E$, $F/D$, $B\flat/G$, $E\flat/C$, and $A\flat/F$ major/minor scales.}\label{fig8}
\end{figure}

\begin{figure}[t]
\centering
\includegraphics[width=0.7\textwidth]{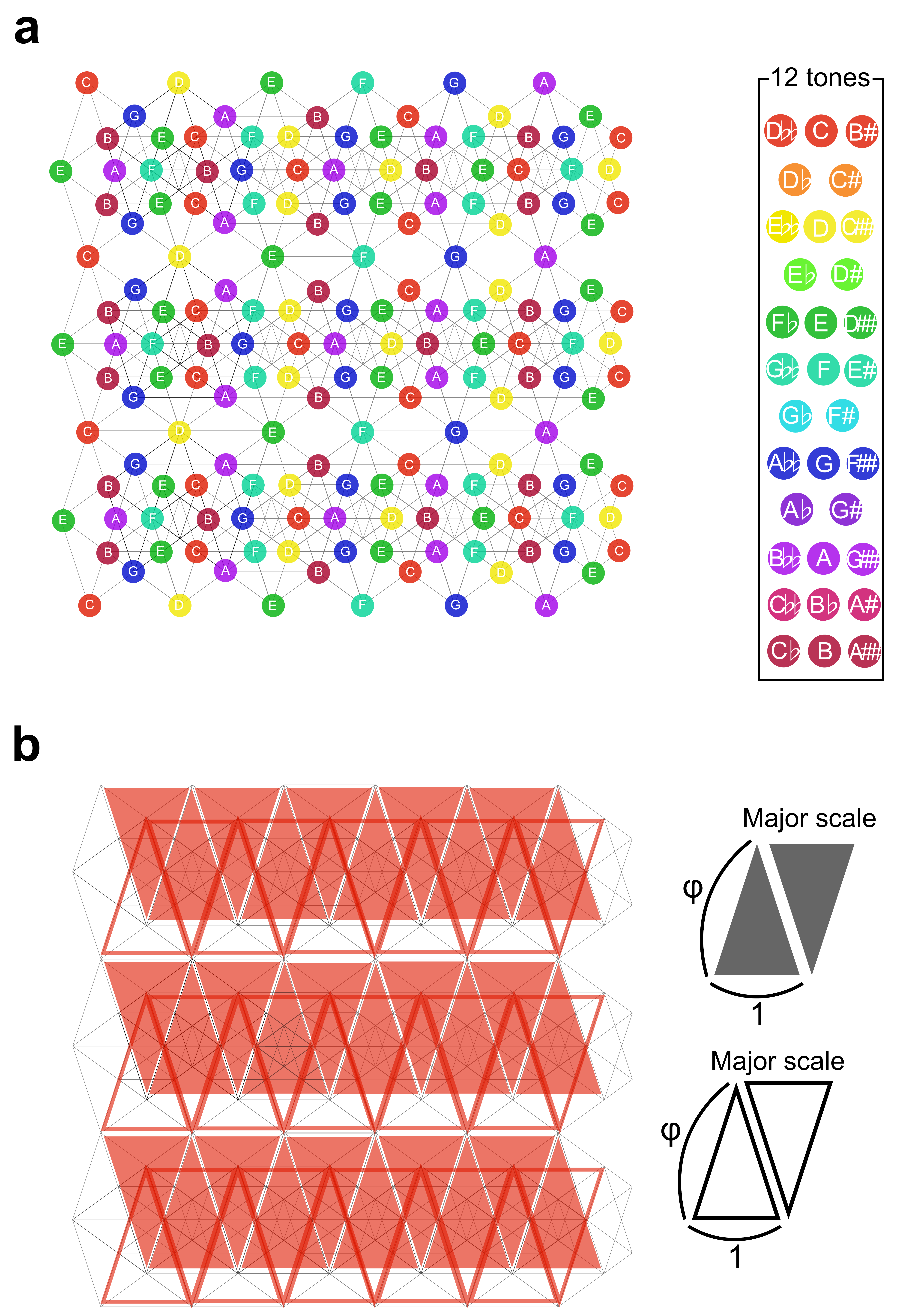}
\caption{\textbf{a}, A tone network obtained by the horizontal extension shown in Fig.~\ref{fig7}\textbf{b} and vertical extension shown in Fig.~\ref{fig8}\textbf{b}. \textbf{b}, The $C$ major scale is represented by the golden triangles and the major scales other than the $C$ major scale cannot be represented by a golden triangle.}\label{fig9}
\end{figure}

\begin{figure}[t]
\centering
\includegraphics[width=0.8\textwidth]{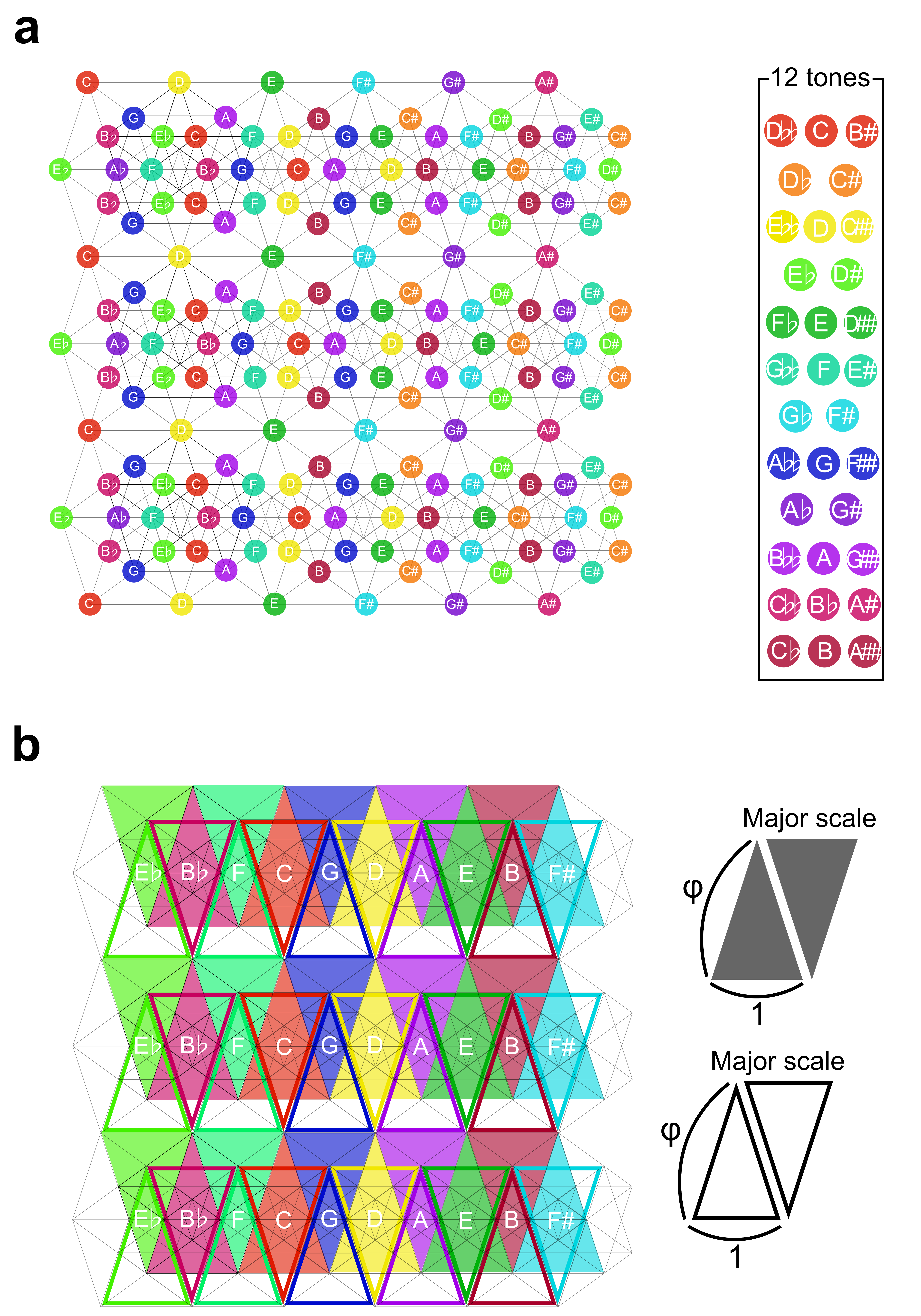}
\caption{\textbf{a}, A tone networks obtained by the horizontal extension shown in Fig.~\ref{fig7}\textbf{a} and vertical extension shown in Fig.~\ref{fig8}\textbf{b}. \textbf{b}, Any major/minor scales can/cannot be represented by a golden triangle.}\label{fig10}
\end{figure}

In addition, the golden Tonnetz can represent the Gregorian modes.
For example, the red lines on the top left figure in Fig.~\ref{fig4}a show that the neighboring tones in the Lydian mode rooted on $C$ neighbor in terms of the graph topology.
Similarly, one can show that all the Gregorian (Lydian, Ionian, Mixolydian, Dorian, Aeolian, Phrygian, and Locrian) modes rooted on any tone can be represented by topologically connected figures in the golden Tonnetz.
As a result, one can represent all the natural scales and triads and Gregorian modes by topologically connected figures in the golden Tonnetz.

\begin{figure}[t]
\centering
\includegraphics[width=\textwidth]{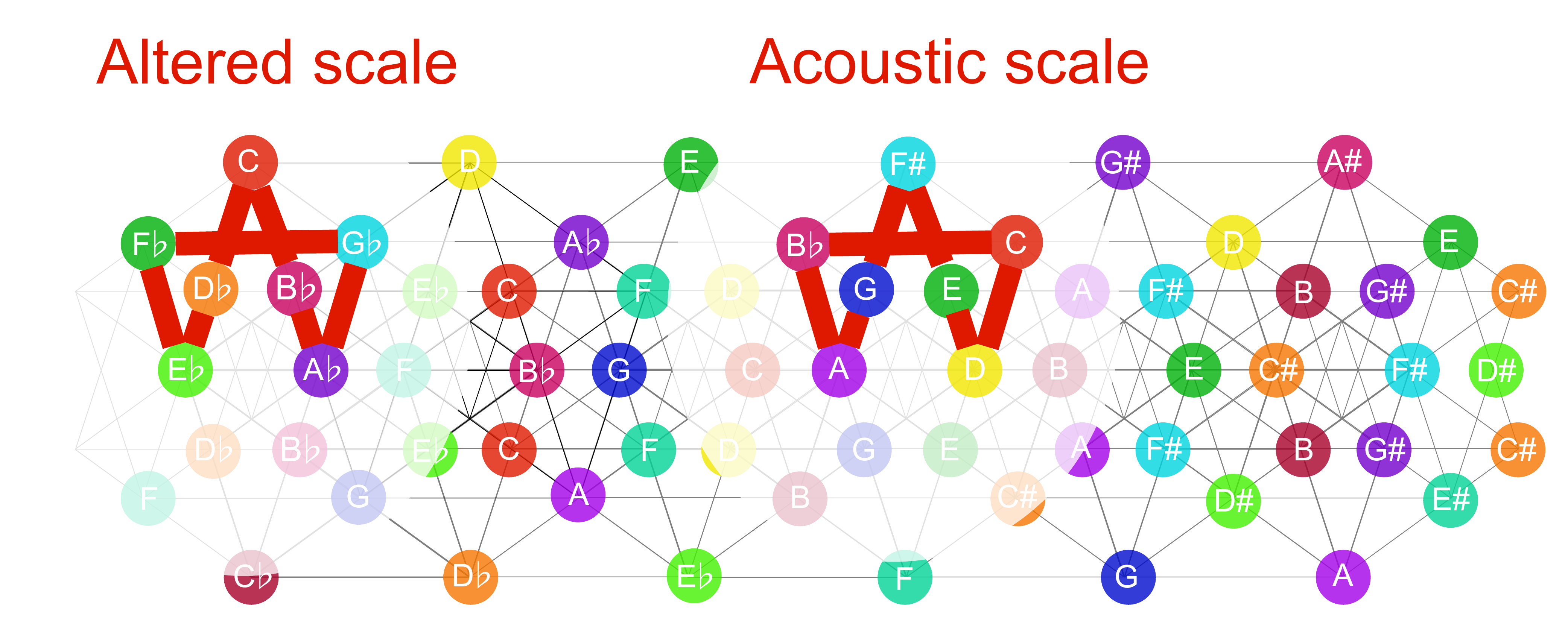}
\caption{Representation of acoustic and altered scales by topologically connected figures on the golden Tonnetz.}\label{fig11}
\end{figure}

The topologically connected figures in the golden Tonnetz can represent other musical scales such as acoustic and altered scales (Fig.~\ref{fig11}).

\section{Conclusion}\label{sec13}
We proposed the concept of golden Tonnetz that extends the concept of Tonnetz by using golden triangles and gnomons.
The golden Tonnetz has a structure in which moving horizontally corresponds to ascending by a perfect fifth, and moving vertically corresponds to ascending by a semitone.
The golden Tonnetz involves not only the relative, parallel, and leading-tone exchange transformation in Neo-Riemannian theory with golden triangles and gnomons but also represents all the major/minor scales and Gregorian modes by topologically connected figures in the golden Tonnetz.
Especially, all the major/minor triads and scales can be represented by golden triangles or gnomons in the golden Tonnetz.

\backmatter





\begin{appendices}
\end{appendices}


\bibliography{sn-bibliography}

\end{document}